\documentclass[twocolumn,trackchanges]{aastex62}

\usepackage{amsmath}	% Advanced maths commands
\definecolor{red}{RGB}{190,0,0}
\definecolor{pink}{RGB}{255,51,153}
\definecolor{green}{RGB}{0,153,0}

\definecolor{blue}{RGB}{20,20,255}

\definecolor{orange}{RGB}{255,165,0}

\definecolor{cyan}{RGB}{50,190,190}

\newcommand{\msun}{\mbox{$M_{\odot}$}}
\newcommand{\msunyr}{\mbox{$M_{\odot}\,{\rm yr^{-1}}$}}
\newcommand{\kms}{\mbox{${\rm km\,s^{-1}}$}}

\newcommand{\cmq}{\mbox{$\,{\rm cm^{-3}}$}}
\newcommand{\nH}{\mbox{$n_{\rm H}$}}

\newcommand{\Lya}{\mbox{$\rm Ly\alpha$}}

\graphicspath{{./}{figures/}}

\received{\today}
\shorttitle{Impact of radiation feedback on GC candidates}
\shortauthors{Daniel Han et al.}

\begin{document}

\title{Impact of Radiation Feedback on the Formation of Globular Cluster Candidates during Cloud--Cloud Collisions}

%% The new \altaffiliation can be used to indicate some secondary information
%% such as fellowships. This command produces a non-numeric footnote that is
%% set away from the numeric \affiliation footnotes.  NOTE that if an
%% \altaffiliation command is used it must come BEFORE the \affiliation call,
%% right after the \author command, in order to place the footnotes in
%% the proper location.
%%
%% Use \email to set provide email addresses. Each \email will appear on its
%% own line so you can put multiple email address in one \email call. A new
%% \correspondingauthor command is available in V6.2 to identify the
%% corresponding author of the manuscript. It is the author's responsibility
%% to make sure this name is also in the author list.

%\correspondingauthor{August Muench}
%\email{greg.schwarz@aas.org, gus.muench@aas.org}

\author[0000-0002-2624-3129]{Daniel Han}
\email{daniel.han@yonsei.ac.kr}
\affiliation{Department of Astronomy, Yonsei University, 50 Yonsei-ro, Seodaemun-gu, Seoul 03722, Republic of Korea}

\author[0000-0002-3950-3997]{Taysun Kimm}
\email{tkimm@yonsei.ac.kr}
\affiliation{Department of Astronomy, Yonsei University, 50 Yonsei-ro, Seodaemun-gu, Seoul 03722, Republic of Korea}

\author{Harley Katz}
\affiliation{Astrophysics, University of Oxford, Denys Wilkinson Building, Keble Road, Oxford OX1 3RH, UK}

\author[0000-0002-8140-0422]{Julien Devriendt}
\affiliation{Astrophysics, University of Oxford, Denys Wilkinson Building, Keble Road, Oxford OX1 3RH, UK}

\author{Adrianne Slyz}
\affiliation{Astrophysics, University of Oxford, Denys Wilkinson Building, Keble Road, Oxford OX1 3RH, UK}

\begin{abstract}

To understand the impact of radiation feedback during the formation of a globular  cluster (GC), we simulate a head-on collision of two turbulent giant molecular clouds (GMCs).
A series of idealized radiation-hydrodynamic simulations is performed, with and without stellar radiation or Type II supernovae.
We find that a gravitationally bound, compact star cluster of mass $M_{\rm GC} \sim 10^5\,\msun$ forms within $\approx 3\,{\rm Myr}$ when two GMCs with mass $M_{\rm GMC}=3.6\times10^5\,\msun$ collide.
The GC candidate does not form during a single collapsing event but emerges due to the mergers of local dense gas clumps and gas accretion.
The momentum transfer due to the absorption of the ionizing radiation is the dominant feedback process that suppresses the gas collapse and photoionization becomes efficient once a sufficient number of stars form. 
The cluster mass is larger by a factor of $\sim 2$ when the radiation feedback is neglected, and the difference is slightly more pronounced (16\%) when extreme \Lya\ feedback is considered in the fiducial run.
In the simulations with radiation feedback, supernovae explode after the star-forming clouds are dispersed, and their metal ejecta are not instantaneously recycled to form stars.
\end{abstract}


\section{Introduction} \label{sec:intro}

The formation of a globular cluster (GC) has been a long-standing puzzle in galactic astronomy. 
\citet{peebles1984} first put forward the idea that the primordial gas in a dark matter halo may rapidly condense into a dense structure due to Lyman alpha cooling.
Cosmological radiation-hydrodynamic (RHD) simulations verified that halos with mass $M_h\sim10^8\,\msun$ can form a dense cluster when purely atomic gas collapses during the transition from a molecular-cooling to atomic-cooling regime \citep{kimm2016}; however, the predicted internal metallicity spread in each GC candidate was significantly larger than that observed.
Moreover, \citet{trenti2015} proposed a similar scenario wherein the mergers of star-free mini-halos cause shock-induced gas collapse but the halo gas needs to be somehow pre-enriched by external sources to match the observed GC metallicity.
Alternatively, a rapid collapse may occur inside a massive giant molecular cloud (GMC) in star-forming galaxies at high redshift \citep[][]{kravtsov2005,howard2019}.
This idea forms the basis of tagging methods wherein a fraction of star particles are assumed to represent potential GC candidates \citep[e.g.][]{renaud2017,pfeffer2018}.
Despite their simplicity, the models can successfully reproduce the color bimodality in GCs, which makes the formation channel appealing. 
The most relevant observational evidence in support of this scenario stems from the detection of young super star clusters in the Antennae galaxies \citep{whitmore1995}.
Additionally, compressive tides and/or shock compression due to galaxy mergers are likely to have played an important role in GC formation \citep{renaud2015,kimjh2018,lahen2019}.
Improving upon tagging methods, \citet{grudic2022b} applied a cluster formation model derived from their GMC simulations \citep{grudic2021} to a Milky Way-like galaxy, and showed that the cluster mass--size relation observed in nearby star-forming galaxies can be explained if denser GMCs form clusters with high efficiency.
However, the detailed formation processes have not yet been rigorously investigated.

\begin{table*}[ht!]
	\centering
	\caption{Input parameters and physical ingredients of the simulations. From left to right, each column indicates the initial gas mass of a cloud ($M_{\rm GMC}$), minimum cell size ($\Delta x_{\rm min}$), mass resolution of a star particle ($m_{\rm star}^{\rm res}$), inclusion of radiation feedback (RF), SN, and \Lya\ pressure, time at the end of simulation ($t_{\rm final}$), total mass of star particles ($M_{\rm star}$), the half-mass radius ($R_{\rm c,h}$), and stellar mass of the GC candidate ($M_{\rm GC}$). The length of the simulated box is 512 pc. The simulations with RF include photoionization, direct radiation pressure, and non-thermal pressure due to the multiple scattering of IR photons.}
	\label{tab:tab1}
	\begin{tabular}{lccccccccccl}
		\hline
		\hline
        Name & $M_{\rm GMC}$ &$\Delta x_{\rm min}$  & $m_{\rm    star}^{\rm res}$ & RF & SN & Ly$\alpha$ & $t_{\rm final}$ & $M_{\rm star}$ & $R_{\rm c,h}$ & $M_{\rm GC}$ & Remark \\
        
         & [\msun] & [pc] & [\msun] &  & & & [Myr] & [\msun] & [pc] & [\msun] & \\
        \hline
        \texttt{Run-Fid} & $3.6\times10^5$ & $0.25$ &  $100$ & \checkmark & \checkmark & --  & $13.0$ & $1.446\times10^5$ & $0.33$ & $1.083\times10^5$ & Fiducial \\
        \texttt{Run-Lya} & $3.6\times10^5$ & $0.25$ &  $100$  & \checkmark & \checkmark & \checkmark & $13.0$ & $1.397\times10^5$ & $0.30$ & $0.908\times10^5$ \\
        \texttt{Run-woRF} & $3.6\times10^5$ & $0.25$ &  $100$  & -- & \checkmark & -- & $13.0$ & $3.239 \times10^5$ & $0.35$ & $2.240 \times10^5$\\
        \texttt{Run-HR} & $3.6\times10^5$ & $0.125$ & $100$ & \checkmark & \checkmark & --  & $10.6$ & $1.303\times10^5$ & $0.23$ & $1.068\times10^5$ \\
        \texttt{Run-Static} & $3.6\times10^5$ & $0.25$ & $100$ & \checkmark & \checkmark & --  & $10.7$ & $0.617\times10^5$ & -- & -- & No collision \\
        \texttt{Run-Fid-woSN} & $3.6\times10^5$ & $0.25$ &  $100$ & \checkmark & -- & --  & $13.0$ & $1.449\times10^5$ & $0.33$ &
        $1.075\times10^5$ \\
        \texttt{Run-Lya-woSN} & $3.6\times10^5$ & $0.25$ &  $100$ & \checkmark & -- & \checkmark  & $13.0$ & $1.419\times10^5$ & $0.34$ &
        $0.871\times10^5$ \\
        \texttt{Run-woRF-woSN} & $3.6\times10^5$ & $0.25$ &  $100$  & -- & -- & -- & $13.9$ & $4.095 \times10^5$ & $0.36$ & $2.716\times10^5$\\
		\hline
	\end{tabular}
\end{table*}

In this paper, we focus on another scenario in which GCs form through the collision of GMCs.
GMCs are known to be birthplaces of present-day star clusters  \citep[e.g.][]{cambresy2002,heyer2006}.
Although the star clusters tend to be less massive and compact than GCs, observational analyses of DR 21 \citep{dobashi2019}, R136 \citep{fukui2017}, and the Antennae galaxies \citep{tsuge2021} suggest that cloud collisions likely trigger star formation.
Indeed, previous studies based on hydrodynamic simulations demonstrated that dense cores form in converging flows \citep{heitsch2009} and perhaps more effectively via cloud--cloud collisions \citep[e.g.,][]{takahira2018,sakre2021}.
Furthermore, \citet{tanvir2020}  showed that considerably more stars are created when colliding clouds are gravitationally bound, than when bound--unbound or unbound--unbound clouds collide.
Moreover, \citet{dobbs2020} demonstrated that for the formation of young massive clusters, high gas densities ($\geq 100\,\cmq$) and velocities greater than $20\,\kms$ are required when two super-virial clouds collide.

Studies have also shown that the effects of radiation need to be understood to comprehend the evolution of GMCs.
\citet{krumholz2007} performed AU-scale simulations of turbulent protostellar cores with mass of $\sim 100\, \msun$ and showed that the initial stellar mass function is significantly affected by the heating due to photoionization.
Additionally, \citet{geen2017} argued that the observed star formation efficiency of  $\sim 10$\,\% in the Milky Way molecular clouds could be reproduced by including ionizing radiation in magnetized GMC simulations.
\citet{kimjg2018} showed that photoionization feedback from stars effectively controls star formation efficiency by disrupting the GMCs within $\sim 5\,{\rm Myr}$, as supported by the CO-to-H$\alpha$ analysis of GMCs in star-forming galaxies \citep[e.g.,][]{chevance2022}.
In contrast, stellar winds, which are another form of early feedback, seemingly play a minor role iRn setting the properties of star clusters \citep{dale2013, grudic2021}.

Recently, \citet{dobbs2022} reported the impact of photoionization in a dense region of a spiral galaxy with converging flows.
More specifically, they simulated cluster formation in strongly and moderately converging regions along a spiral arm in the disks of galaxies similar to the Milky Way.
They found that the star formation rate decrease by $\sim20\%$ in the isolated filamentary structures when radiation feedback was included, whereas the effect was negligible in regions with multiple converging filaments.

Building upon previous studies, this work aims to investigate the impact of radiation feedback on the formation process of a massive compact cluster during cloud--cloud collisions, which may be seen as a progenitor of present-day GCs.
Although such collisions between massive clouds seem to be rare in the local Universe, these may occur more frequently at high redshifts during the formation of dwarf galaxies \citep{kimjh2018,sameie2022}.
Thus, we employ RHD simulations with various forms of feedback, including photoionization, direct radiation pressure, non-thermal pressure due to multiple scattering of IR and \Lya\ photons, and Type II SN explosions.

This paper is organized as follows.
Section 2 describes the initial conditions, numerical methods, and physical parameters of our simulations.
Section 3 examines the formation process of a massive compact cluster and effects of stellar feedback on it.
Section 4 discusses the relative importance of several forms of stellar feedback, the compactness of the massive cluster, and caveats in our simulations.
Finally, Section 5 summarizes our results.

\section{Radiation-Hydrodynamic Simulations}
\label{sec:setup}

We perform RHD simulations of cloud--cloud collisions with the adaptive mesh refinement code {\sc ramses-rt} \citep{teyssier2002, rosdahl2015}.
For each spherical cloud, an initial number density of $\nH = 100\,\cmq$ and temperature of $T = 100\,K$ are adopted.
The initial radius of the clouds is set as $30\, {\rm pc}$, based on the mass--size relations of observed GMCs \citep{roman2010, heyer2009, oka2001}.
We place two clouds with mass $M_{\rm GMC}=3.6\times10^5\,\msun\,$ 100 pc apart in a uniform medium with density $\nH=0.54\,\cmq$ and temperature $10^4\,{\rm K}$.
Turbulent velocity fields following the Kolmogorov power spectrum are developed over half of the free-fall time ($0.5\,t_{\rm ff}\approx 2.2\,{\rm Myr}$) of the cloud, at which point the virial parameter of each cloud reaches $\alpha_{\rm vir}=0.7$.
We generate the random force fields via an Ornstein-Uhlenbeck process in a Fourier domain with a compressive to solenoidal ratio of 1:3 \citep[see e.g.,][]{federrath2010}.
To avoid a bulk cloud-scale motion due to the random forces, we limit the wave numbers to be greater than 6.
The simulated volume of $(512\,{\rm pc})^3$ is covered by $128^3$ coarse cells with 4 pc resolution, which are further refined up to 0.25 pc (fiducial) or 0.125 pc (high-resolution case) when the gas mass in a cell exceeds $0.1\,\msun$.
To better capture turbulent structures, the thermal Jeans length is resolved using at least 32 cells until the maximum refinement level is reached \citep{federrath2012}.
Additionally, cells are refined if the change in pressure with respect to the neighboring cells is greater than 100\%. To reduce the computational cost, we limit the refinement to regions representing a GMC using two passive scalars.
Specifically, the outside of the clouds is minimally refined unless $\ge 1\%$ of the cell mass stems from either of the clouds. 
These yield $\sim 10^8 - 2\times 10^8$ computational cells, of which $\sim$ 1/3 are resolved at the maximum level.

To simulate star formation, the sink particle algorithm developed by \citet{bleuler2014} is employed.
First, we identify gravitationally bound gas clumps using the Parallel HiErarchical Watershed ({\sc phew}) algorithm \citep{bleuler2015} with a density of $0.1\,\rho_{\rm th}$, where $\rho_{\rm th}$ is the threshold density for sink particle formation.
Following \citet{bleuler2014}, we adopt $\rho_{\rm th}=\pi c_s ^2/16 G \Delta x_{\rm min}^2$, where $c_s$ is the speed of sound, $G$ is the gravitational constant, and $\Delta x_{\rm min}$ is the size of the finest cell.
The corresponding $\rho_{\rm th}$ values for the runs with $\Delta x_{\rm min}=0.25\,{\rm pc}$ and $0.125\,{\rm pc}$ resolutions are $3.0\times10^{-20}\,{\rm g\,cm^{-3}}$ and $1.2\times10^{-19}\,{\rm g\,cm^{-3}}$, respectively.
The accretion onto the sink particle is modeled using a flux accretion scheme.
If the mass of a sink particle exceeds $m_{\rm star}^{\rm res}=100\,\msun$, a star particle of mass $m_{\rm star}^{\rm res}$ is created, which generates ultraviolet and optical photons, as detailed below. 

The numerical methods and physical models of gas cooling and stellar feedback employed herein are identical to those employed in \citet{kimm2019}.
The hydrodynamic equations are solved using the MUSCL scheme with a multi-dimensional MonCen limiter and HLLC Riemann solver \citep{toro1994}.
For radiative transfer, a GLF solver is adopted with a reduced speed of light approximation ($10^{-3}\,c$, \citealt{rosdahl2013}).
The Courant number of 0.8 is used.
To model photoionization, direct radiation pressure, non-thermal pressure due to multiple scattering of IR photons, and photo-electric heating due to dust, eight photon groups are employed, as detailed in \citet[][see their Table 1]{kimm2019} in units of eV, these are: [0.1, 1.0), [1.0, 5.6), [5.6, 11.2), [11.2, 13.6), [13.6, 15.2), [15.2, 24.59), [24.59, 54.42), and [54.42,$\infty$).
The explosion of Type II SNe is modeled using a mechanical scheme developed by \citet{kimm2015}.
To bracket the impact of strong radiation feedback, we also examine an extreme scenario where additional pressure due to \Lya\ photons is included, assuming that the local velocity structures of the scattering medium are static \citep{kimm2018}.
The spectral energy distribution of each star particle is derived from the Binary Population And Spectral Synthesis model (\citealt{stanway2016}, BPASS v2.0) assuming a Kroupa initial mass function (IMF) with lower and upper cut-off masses of $0.1$ and $100\,\msun$, respectively.
The initial metallicity of the clouds is set as $Z_{\rm ini} = 0.1\,Z_\odot$, where $Z_\odot=0.02$ is the solar metallicity.
We adopt the metal yield of 0.075 for SN explosions.
Gas cooling due to primordial species (HI, HII, HeI, HeII, HeIII, ${\rm H_2}$, and e$^-$) is included by solving the non-equilibrium chemistry \citep{rosdahl2013,katz2017}, and the cooling due to metallic species is included based on the Cloudy calculations \citep{ferland1998} and fine-structure line cooling by \citet[][`cc07' model in {\sc ramses}]{rosen1995}.

\begin{figure}
\centering
\includegraphics[width=\linewidth]{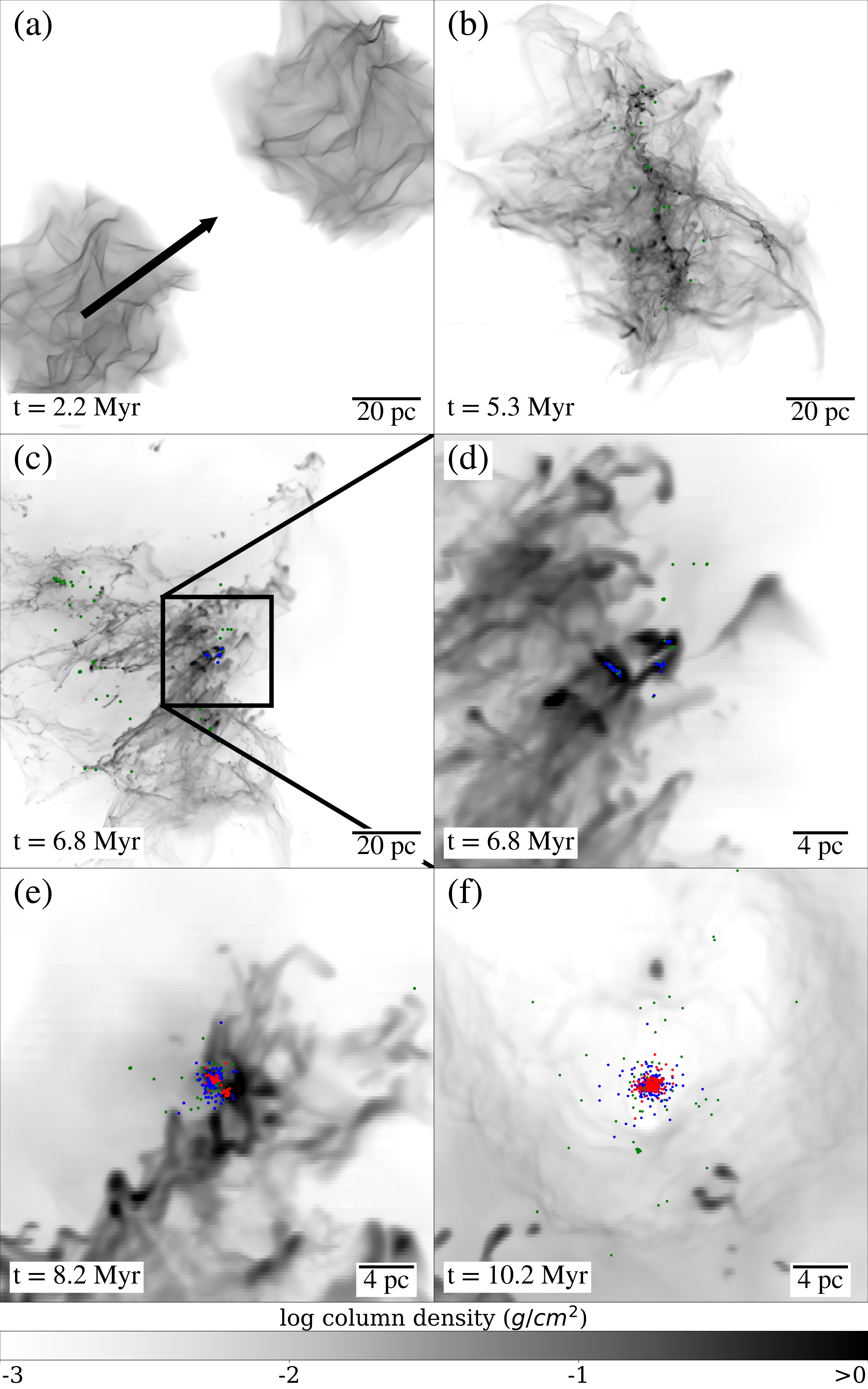}
\caption{Formation of a massive compact cluster (i.e. potential GC) in the fiducial run (\texttt{Run-Fid}). Each panel displays the projected distributions of gas densities in the $(128\,{\rm pc})^3$ (panels (a)--(c)) or $(32\,{\rm pc})^3$ cube (panels (d)--(f)). The blue and red dots represent the stars formed during the first and second star formation events, respectively, whereas the green dots represent stars that do not belong to the  massive cluster. Panel (a) shows the initial turbulent structures of the two GMCs with mass $3.6\times10^5\,\msun$. Different panels correspond to the moment at which the GMCs coalesce (b), the moment at which half of the population \texttt{A} (i.e., pre-merger of the gas clumps) (d) or population \texttt{B} (i.e., post-merger) emerges (e), and the moment at which the GMCs are destroyed by the radiation feedback (f). The gray colors indicate the gas column density, as shown in the legend. }
\label{fig:fig_fiducial}
\end{figure}

To model the head-on collision of two massive clouds, the velocity $v_{\rm col}$ is added to the cloud in the lower left at $t=0.5\,t_{\rm ff}$ (Figure~\ref{fig:fig_fiducial}, panel-(a)), and the other is kept stationary to minimize the numerical diffusion caused by the relative motion of clouds across the hydrodynamic grid \citep[e.g.][]{robertson2010}.
In this study, $v_{\rm col}$ is set to $30 \,\kms$, motivated by the velocity range of collisions that induce locally enhanced star formation according to the recent idealized simulations of a disk galaxy embedded in a $10^{11}\,\msun$ dark matter halo \citep[][priv.comm.]{yoo2020}\footnote{We also examined the case with $v_{\rm col}=100\,\kms$, motivated by the high-velocity collision inferred from the radio observation of the formation of super star clusters in the Antennae galaxies \citep{tsuge2021}. 
However, we found that with $v_{\rm col}=100\,\kms$ the clouds become gravitationally unbound and stretched along the direction of the collision, they form no star clusters.}. 
We also perform a simulation with no cloud collision (\texttt{Run-Static}, $v_{\rm col} = 0$) as a control sample.
Each simulation is run for $11$--$14\,{\rm Myr}$, at which point no additional stars are formed during the last $\ge 1$ Myr.
The star clusters are identified using the \texttt{DBSCAN} method \citep{ester1996}.
The final cluster mass $M_{\rm GC}$ and radius $R_{\rm GC,h}$ are measured 1 Myr after the end of star formation in the GC candidates, except in the \texttt{Run-woRF-woSN} run where star formation continues.
These correspond to 10.7, 10.1, 12.3, 10.6, 11.0, and 12.6 Myr for \texttt{Run-Fid}, \texttt{Run-Lya}, \texttt{Run-woRF}, \texttt{Run-HR}, \texttt{Run-Fid-woSN}, and \texttt{Run-Lya-woSN} run, respectively. In the case of \texttt{Run-Static} where no GC candidate is formed, we measure the stellar mass at $t = 10.7\, {\rm Myr}$, which is 1 Myr after the end of star formation in the entire simulated volume.
Table~\ref{tab:tab1} summarizes the setup of our simulations and physical properties of the GC candidate.

%=============================================
\section{Results} \label{sec:results}
%=============================================

\subsection{Formation of a massive compact cluster}

\begin{figure}
\centering
\includegraphics[width=\linewidth]{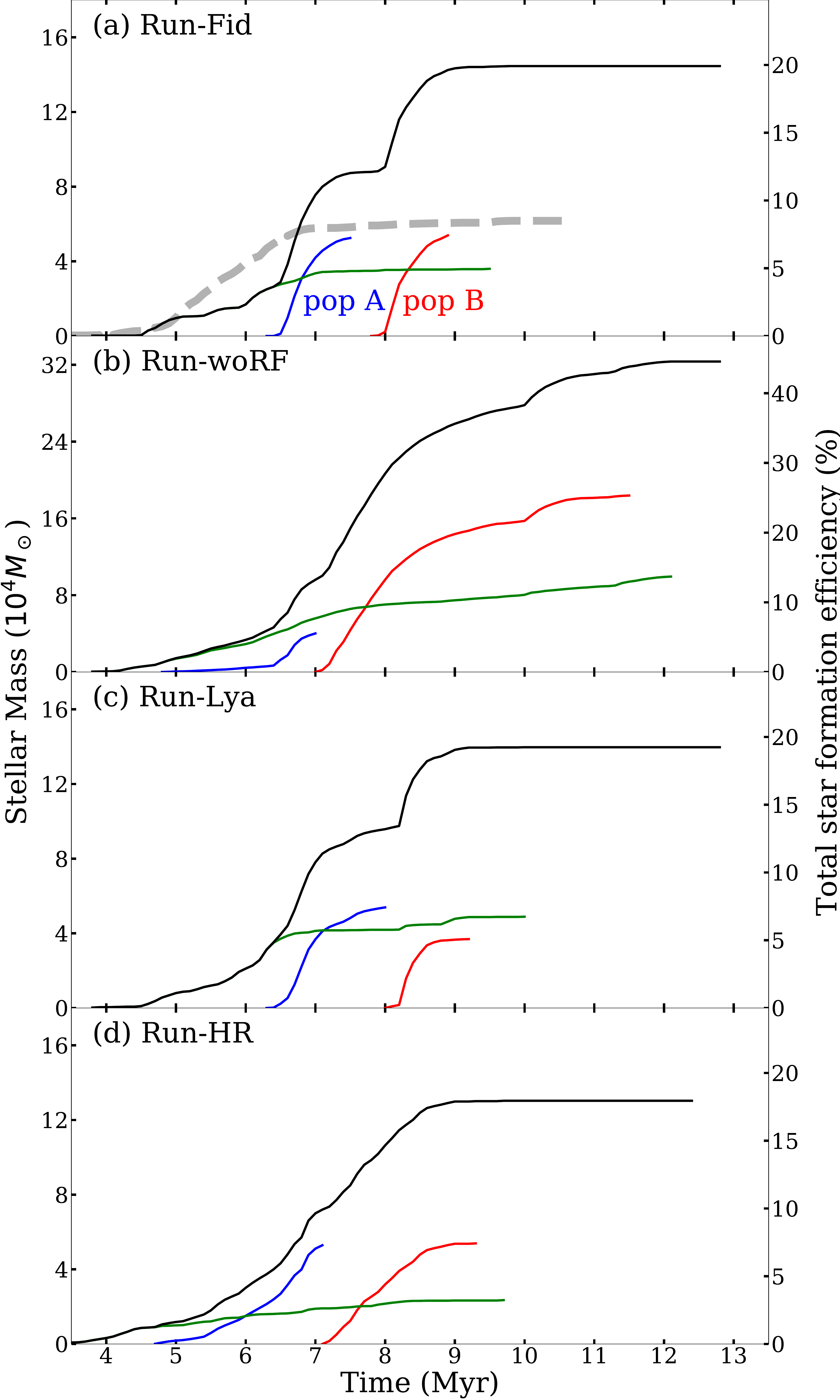}
\caption{Star formation histories of the colliding GMCs. The blue and red lines denote the pre-merger (\texttt{A}) and post-merger (\texttt{B}) star formation episodes of the GC candidate, respectively. The black and green lines represent the stellar mass growth of the total and non-GC populations, respectively. The color schemes are the same as in Figure~\ref{fig:fig_fiducial}. Stellar mass is indicated on the left axis whereas the total star formation efficiencies ($M_{\rm star}/2M_{\rm GMC}$) are shown on the right axis. The grey dashed line in panel (a) indicates the total stellar mass growth in \texttt{Run-Static}.}
\label{fig:fig_SFH}
\end{figure}

\begin{figure*}
\centering
\includegraphics[width=\linewidth]{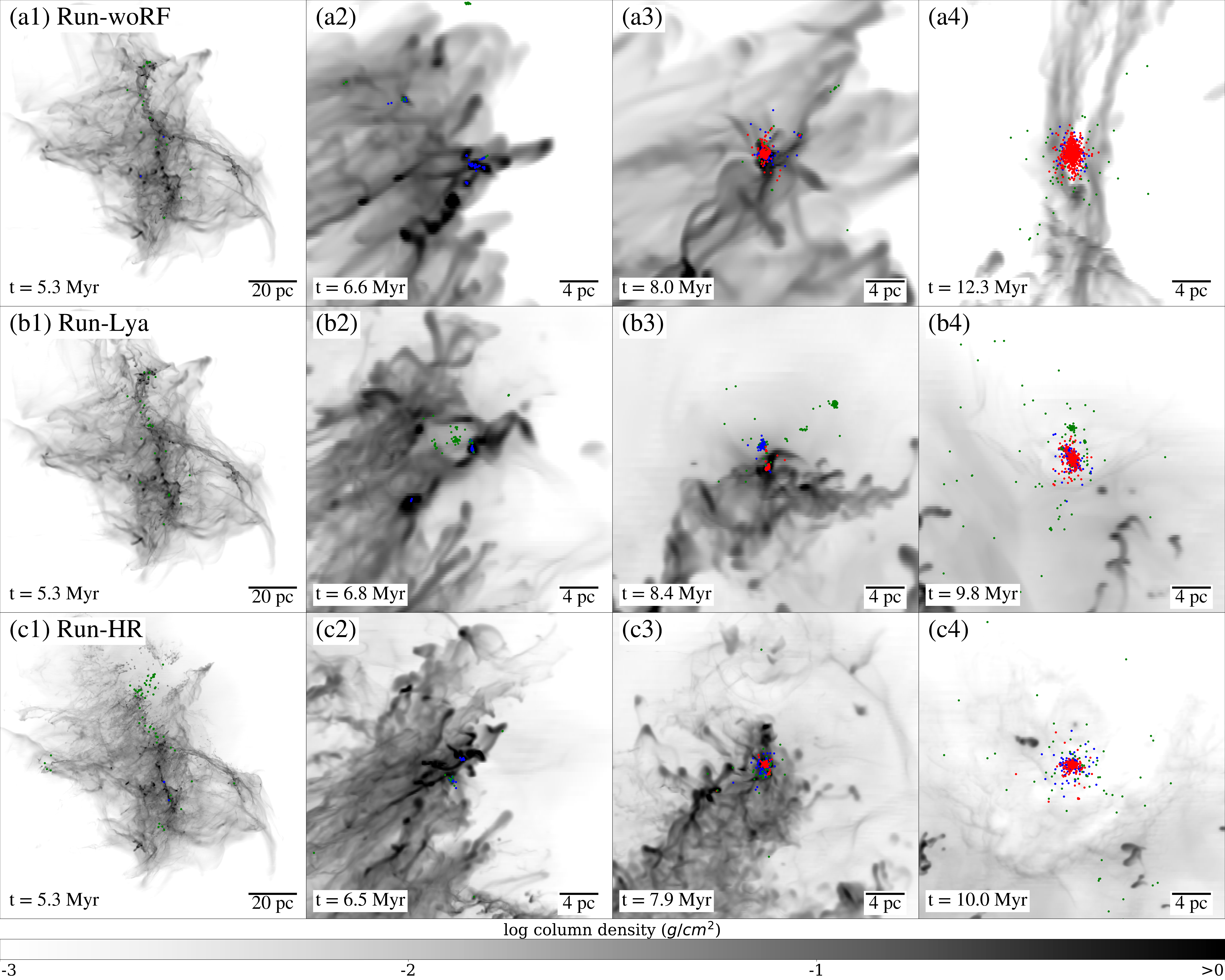}
\caption{Formation of a massive compact cluster in the run without radiation feedback (\texttt{Run-woRF}; top panels), with Ly$\alpha$ feedback (\texttt{Run-Lya}; middle panels), and with higher resolution (\texttt{Run-HR}; bottom panels). Each panel exhibits the projected distributions of gas densities in the $(128\,{\rm pc})^3$ (the first column) and $(32\,{\rm pc})^3$ cubes (the other three columns). From left to right, the panels show the coalescence of two GMCs, the first or second star formation episodes,  and the massive gas expulsion by radiation (panels (b4) and (c4)) or supernova feedback (panel (a4)). The colored dots represent star particles formed in different phases, as in Figure~\ref{fig:fig_fiducial}.
}
\label{fig:img_else}
\end{figure*}

We begin by showing how a compact, massive star cluster forms when two GMCs, each of mass $M_{\rm GMC} = 3.6\times 10^5\, \msun$, collide head-on with $v_{\rm col}=30\,\kms$ (\texttt{Run-Fid}).
As shown in Figure~\ref{fig:fig_fiducial}, the two clouds first coalesce at $t = 5.3\,{\rm Myr}$ with a mild compression at their interface (panel-b).
Because of the turbulent structures of the GMCs, star formation is not immediately triggered at the interface \citep[c.f.,][]{hunter2021}, but the gas continues to collapse locally.
At this stage, stars with the total stellar mass of $1.5\times 10^4\,\msun$ are formed; they account for only $\sim 10\,\%$ of the total stellar mass produced by the end of the simulation.
Note that none of these stars belong to the massive compact star cluster that emerges later.
Most of the potential GC population starts appearing $1\,{\rm Myr}$ after the coalescence ($t>6.5\,{\rm Myr}$).
Several dense gas clumps that are located within a small region ($\la 10\,{\rm pc}$) first form stars individually and then start to merge around $t=7.8\,{\rm Myr}$ (not shown, but see the blue dots in Figure~\ref{fig:fig_fiducial}).
The star cluster formed from the multiple clumps is similar to the GC formation picture presented in the hydrodynamic simulations of dwarf galaxy mergers \citep{lahen2019}.
Neighboring clouds and diffuse gas continue to accrete onto the cluster and form stars (red dots) until the radiation feedback entirely disrupts the gas clump at $t\approx10\,{\rm Myr}$.
This is corroborated in Figure~\ref{fig:fig_SFH} (panel-a), which depicts the star formation histories of the colliding GMCs.
In this figure, the two populations, shown using blue and red lines (\texttt{A} and \texttt{B}, respectively), indicate that the compact cluster does not form in a single event but emerges due to the local collapse and merger, followed by gas accretion.
The formation process occurs over a short timescale of $\approx 2$--$4\, {\rm Myr}$, resulting in a final cluster of mass $M_{\rm GC}=1.08\times10^5\,\msun$ with a half-mass radius of 0.33 pc in the fiducial case.

We note that the star formation efficiency per free-fall time is considerably higher ($\sim 20\%$) than in observed molecular clouds (1--2\%) \citep{krumholz2007b, utomo2018}. It is generally argued that this observed low efficiency arises from turbulence-regulated star formation in clouds with relatively low column densities $\sim 100\,\msun {\rm pc}^{-2}$ and marginally high virial parameters $\alpha_{\rm vir}\sim 1$--$2$ \citep[e.g.,][]{krumholz2006, krumholz2009}. In contrast, our simulated GMCs have lower virial parameters ($\alpha_{\rm vir} = 0.7$) and higher gas column densities ($\approx 300\,\msun {\rm pc}^{-2}$), compared to GMCs located in the Local Group galaxies \citep{bolatto2008} or the Milky Way \citep{heyer2009}. Indeed, in model \texttt{Run-static} where no collision is introduced, $\approx 8\%$ of gas is converted into stars per free-fall time (Figure~\ref{fig:fig_SFH} (panel-a), grey dashed line), although no GC candidate is formed in \texttt{Run-static}. Such a high star formation efficiency is often obtained in the simulations of GMCs with high gas surface densities \citep[e.g.,][]{grudic2021,kimm2022}. We surmise that these more extreme physical conditions, typical of high-redshift galaxies, increase the likelihood of forming dense and compact clumps during cloud--cloud collisions, thus providing a favorable environment for viable GC candidate formation by boosting star formation efficiency.

In addition to the main cluster, the simulated GMCs form small compact star clusters with radius $\la 1\, {\rm pc}$.
In the fiducial run, two stellar aggregations with $>40$ particles are identified; their masses and half-mass radii are $9.6\times10^3\,\msun$, $4.9\times10^3\,\msun$ and $\approx 0.2 \,{\rm pc}$, $0.1\, {\rm pc}$, respectively.
They form either in a single clump (former) or during the merging of gas clumps (latter); however, no subsequent gas accretion occurs in these small systems, unlike the formation process of massive GC candidates. Such stellar aggregations are also formed in the simulation without cloud collision (\texttt{Run-Static}).
We note that these small clusters do not form in the higher resolution run (\texttt{Run-HR}), as gas fragmentation is better captured, yielding even smaller clumps in size and mass \citep[e.g.,][]{price2010,federrath2012}.
Stellar feedback then efficiently disperses gas clumps before they merge and form a massive clump that would otherwise end up as a star cluster of mass $9.6\times10^3\,\msun$.
In contrast, the GC candidate is still present in \texttt{Run-HR}, denoting that the formation of the massive compact cluster is less affected by resolution.

\subsection{Effects of radiation feedback}

With the potential GC, we study the impact of radiation feedback on the stellar mass, gas inflow/outflow, and metal enrichment of the compact clusters.

\subsubsection{Star formation history}

Figures~\ref{fig:fig_SFH}--\ref{fig:img_else} present the star formation histories and column density distributions of the run without radiation feedback (\texttt{Run-woRF})
where the star particles do not produce photons.
Thus, both photoionization and radiation pressure due to UV and the multiple scattering of IR photons are ignored, while still considering SN explosions. 

We find that an approximately twice more massive GC candidate forms in the \texttt{Run-woRF} run than in the fiducial run.
In Figure~\ref{fig:fig_SFH}, the stellar mass of the population \texttt{A} (stars formed before the one final massive clump appears) is slightly reduced, but this is because the mergers between gas clumps take place earlier than those in the fiducial run.
The growth of stellar masses is not controlled in \texttt{Run-woRF}, and stars continue forming even after $t\ga 9\,{\rm Myr}$ at which point the radiation feedback blows the gas away in \texttt{Run-Fid} (see also Figure~\ref{fig:img_else}).
Consequently, 31\% of the two GMCs is converted into the GC candidate within $\approx 6\,{\rm Myr}$ in \texttt{Run-woRF}, while only 15\% forms stars in the \texttt{Run-Fid} run.
Additionally, the mass of the second or third massive stellar aggregations increases by a factor of 2--3 ($3\times10^4$ or $1\times10^4\,\msun$, respectively) in the absence of radiation feedback.
This clearly demonstrates the importance of radiation in the formation of compact clusters.
Thus, any simulation that does not account for radiation feedback will likely overestimate the stellar mass of a GC by a factor of $\sim 2$. 

The predicted stellar mass of a potential GC may decrease even further if additional nonthermal pressure operates on star-forming sites (see the second row in Figure~\ref{fig:img_else}).
As demonstrated by \citet[][see also \citealt{smith2017,tomaselli2021}]{dijkstra2008},  the resonant scattering of \Lya\ photons can accelerate dust-poor gas to several tens of \kms, possibly suppressing the formation of metal-poor star clusters \citep[e.g.,][]{kimm2018,abe2018}.
Since the column density of neutral hydrogen in our metal-poor star-forming clumps tends to be high ($N_{\rm HI}\ga 10^{22}\,{\rm cm^{-2}}$),  \Lya\ photons can easily scatter $10^{7}$--$10^{8}$ times.
Accordingly, the typical force multiplier of \Lya\ photons ($M_F$), which is a measure of the enhancement in momentum transfer compared to the single scattering limit, can be large ($M_F\sim10$--$50$) \citep[][see their Fig. 1]{kimm2018}.
Thus, \Lya\ pressure may play an important role in GC-forming sites.
Strictly speaking, the number of resonant scatterings sensitively depends on the velocity structure of the ISM, but here the scattering medium is assumed to be static to maximize the impact of \Lya\ feedback.
Therefore, the \texttt{Run-Lya} run should be considered as an extreme model where early feedback due to radiation is enhanced, disrupting star-forming clumps.
More realistic simulations will likely yield results between the fiducial and \texttt{Run-Lya} model. 

We find that the additional pressure prevents the gas from accreting onto the formation site and suppresses the total stellar mass growth.
The stellar mass of population \texttt{A} is little affected because the gas clumps rapidly collapse before the \Lya\ pressure becomes effective.
However, \Lya\ feedback from this population reduces the gas mass within the radius of 6 pc (10 pc) from the cluster center at $t = 8\,{\rm Myr}$ from $8.9\times10^4\,\msun$ ($11.4\times10^4\,\msun$) in the fiducial run to $6.4\times10^4\,\msun$ ($8.8\times10^4\,\msun$), respectively.
Moreover, as shown in Figure~\ref{fig:fig_SFH} (panel-(c)), \Lya\ pressure intervenes in the star formation of population \texttt{B} in $\Delta t\approx 0.5\,{\rm Myr}$ (see the red line), which is significantly faster than that for the fiducial case, wherein it takes $\Delta t\approx 1.0\,{\rm Myr}$ to quench star formation.
The mass of population B is reduce by 33\% in \texttt{Run-Lya}, and the total stellar mass of the potential GC decreases by 16\% ($M_{\rm GC}=9.08\times10^4\,\msun$).

\subsubsection{Gas inflow and outflow}

\begin{figure}
\centering
\includegraphics[width=\linewidth]{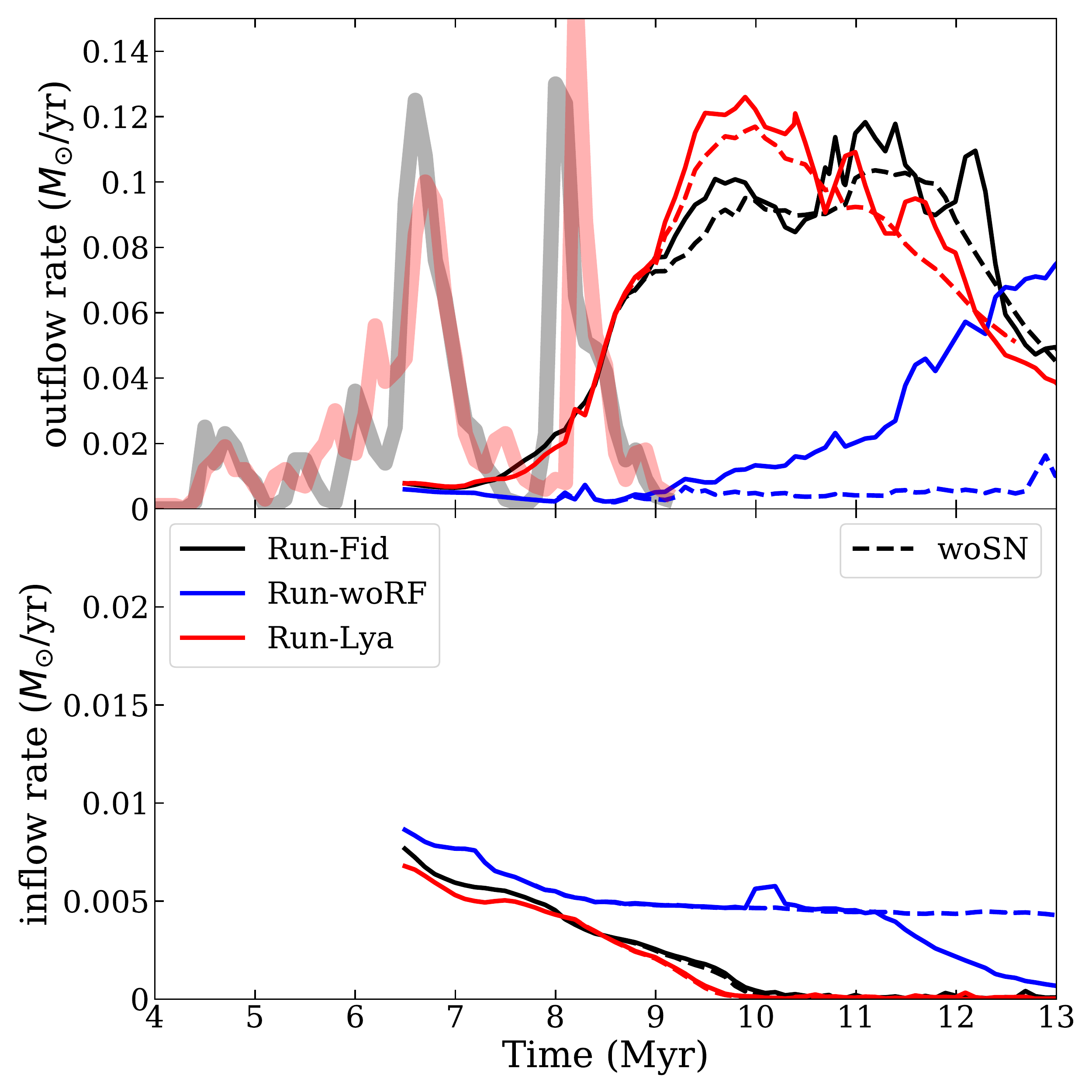}
\caption{Gas outflow (\textit{Top panel}) and inflow (\textit{Bottom Panel}) rates measured at the spherical surface with radius of 128 pc. The center of the sphere at each snapshot is defined as the center of mass of the two GMCs, including stars. The black, blue, and red solid lines represent the \texttt{Run-Fid}, \texttt{Run-woRF} and \texttt{Run-Lya} simulations, respectively. The dashed lines denote the runs without SN explosions. The thick grey and red lines indicate the star formation rates in \texttt{Run-Fid} and \texttt{Run-Lya}, respectively.}
\label{fig:t_flow}
\end{figure}

Figure~\ref{fig:t_flow} presents the gas outflow and inflow rates measured at the spherical surface of radius 128 pc from the center of the two GMCs (gas plus stars). 
The surface is taken as large as possible in order to avoid any confusion due to the colliding motion of the two GMCs. By doing so, we confirm that the net inflow plus  outflow rate is close to zero in \texttt{Run-woRF-woSN}.

Outflows start appearing at $t\sim7\,{\rm Myr}$ even before the massive compact cluster forms and destroys the gas clumps.
This happens because the radiation from the underlying population in the neighboring region also drives the winds (see the green lines in Figure~\ref{fig:fig_SFH}). 
After $t\sim 9\,{\rm Myr}$, SN explosions from the underlying population also play a role in enhancing the gas outflow, as can be inferred from the comparison with the simulations without SNe (dashed lines in Figure~\ref{fig:t_flow}).
The dense gas in the potential GC-forming site is dispersed only at $t\approx 9.5\,{\rm Myr}$, and SN explosions from the GC candidate are mainly responsible for the enhanced outflows at $t\ga 10.5\,{\rm Myr}$, with velocities up to $v\sim 600\,\kms$ at $t = 13.0$\,{\rm Myr}.
Given that the outflow rates are $\dot{M}_{\rm out}\sim 0.1\,\msunyr$ at $t\ga10\,{\rm Myr}$, the bulk mass loading factor may be estimated as $\dot{M}_{\rm out}/\left<{\dot M}_{\rm star}\right>\sim 3.5$, where $\left<\dot M_{\rm star}\right>$ is simply taken as the total stellar mass divided by 5 Myr.
In contrast, if we take the stellar mass of the underlying population, the mass loading factor would increase by a factor of four ($\dot{M}_{\rm out}/\left<\dot M_{\rm star}\right>\sim 13.8$).

The outflow rates in the fiducial run with radiation feedback are systematically larger than those of \texttt{Run-woRF} at $t\la 12\,{\rm Myr}$.
The typical density and velocity of the outflows are $\nH\sim $ 1--2 $\,\cmq$ and $v_{\rm out}\sim$ 30--100 $\,\kms$, respectively.
Note that even in this case with a high star formation efficiency of $\approx 20\%$, the density of the gas outflows on the GMC scale is significantly smaller than the electron density ($n_e\sim 300\,\cmq$) inferred from the metal emission lines in star-forming galaxies at high redshift \citep[e.g.,][]{sanders2016,schreiber2019}.
We attribute this to the fact that photoionization heating overpressurizes the medium, efficiently lowering the density  \citep[e.g.,][]{krumholz2007}; in contrast shock compression due to SNe may play a role on the galactic scale \citep{schreiber2019}.
The outflowing density and velocity in the \texttt{Run-Lya} run further increase due to the destruction of the dense star-forming clumps because of the strong radiation pressure.
This increases  the outflows by $\sim20\%$.
Furthermore, gas inflow rates are suppressed in the runs with radiation feedback, and even become almost zero within $\sim 3\,{\rm Myr}$ from the onset of star formation; this indicates that the impact of radiation is not limited to a small volume of star-forming clumps but extends to a large solid angle.
The outflows become weak rapidly at $t\ga 12\,{\rm Myr}$, as the majority of the GMC gas is blown out and the average density in the sphere with radius of $128\,{\rm pc}$ is reduced ($\nH\approx 2\,\cmq$ at $t=13\,{\rm Myr}$).

Note that the formation of a potential GC ends before their massive stars explode as SNe. This means that radiation can drastically alter the environment into which SNe propagate \citep[e.g.,][]{chevance2022}.
For example, SNe around $t\approx 11\,{\rm Myr}$ emerge at $\nH\sim10^4\,\cmq$ in the absence of radiation feedback (\texttt{Run-woRF}), but they explode at $\nH\sim0.1\,\cmq$ in the fiducial run (see yellow lines in Figure~\ref{fig:r_fbk}).
Given that the final momentum from SNe ($p_{\rm SN}$) is proportional to $n_{\rm H}^{-2/17}$ \citep[e.g.,][]{thornton1998}, the high-density environments denote that $p_{\rm SN}$ from SNe that emerge early in \texttt{Run-woRF} (i.e., before they destroy the star-forming clouds) will likely be underestimated by $\sim \left( 0.1/10^4\right)^{-2/17}\approx 4$, mitigating the effect of SNe on galactic scales.

\subsubsection{Internal metal enrichment}
\label{sec:metal}

\begin{figure}
\centering
\includegraphics[width=\linewidth]{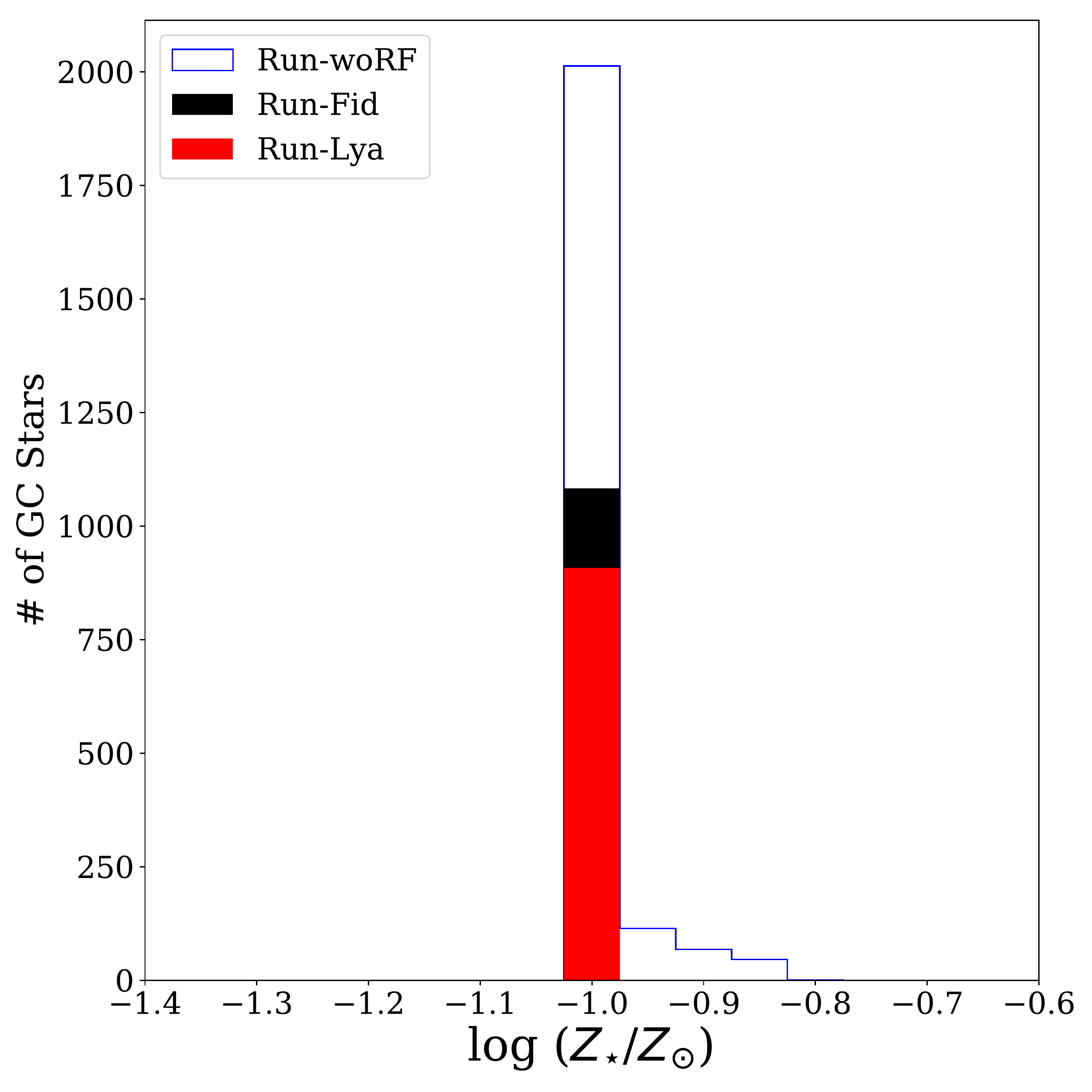}
\caption{Metallicity distribution of the stars in the GC candidate. Different color-codes correspond to different runs, as indicated in the legend. Note that the simulation without radiation feedback over-predicts the internal metal enrichment due to SN explosions. We do not include the distribution from \texttt{Run-HR}, as it is very similar to that from \texttt{Run-Fid}.}
\label{fig:Z_hist}
\end{figure}

As is well established, young GCs are composed of simple stellar populations with a small internal spread in [Fe/H] \citep[e.g.,][]{hollyhead2019}. In general, the observed metallicity spread in Milky Way GCs is $\sim 0.045\,{\rm dex}$ \citep{carretta2009, bailin2019}.
We find that the presence of radiation feedback suppresses the internal metal enrichment due to SNe.
In our simulations, Type II SN explosions occur between 3.5 and 40 Myr when stars more massive than $8\,\msun$ evolve off the main sequence, but the formation of the star clusters ends when their age is less than 3 Myr ($t \approx 9.0\,{\rm Myr}$, Figure~\ref{fig:fig_SFH}).
Thus, no stellar particle is significantly self-enriched by the explosion of SNe inside the GC candidate.
Although some of the stars formed before the collision explode at $t\ga 8.0\,{\rm Myr}$, only a tiny amount of the ejecta is accreted onto the sink and all of the cluster stars exhibit metallicity close to the initial metallicity ($Z_{\rm ini}$), i.e., $\Delta Z_\star / Z_{\rm ini} < 0.01$  (Figure~\ref{fig:Z_hist}). 

In contrast, the run without radiation feedback comprises a non-negligible fraction (16\%) of the GC stars that are self-enriched ($\Delta Z_\star / Z_{\rm ini} > 0.01$).
Owing to the explosions of massive stars and efficient gas consumption, star formation is still quenched quite early ($\sim 6\,{\rm Myr}$) but the metallicity of the cluster stars  increases up to $0.16\,Z_\odot$.
Although this metallicity is not significantly large, care must be taken especially when modeling chemical mixing in multiple populations of GCs using hydrodynamic simulations without radiative transfer.

We note that the presence of radiation feedback does not necessarily indicate zero metallicity spread in GCs.
First, any potential variation in the intrinsic metallicity of the colliding GMCs is neglected in this work \citep[e.g.,][]{de2021}.
If the initial metallicities of the GMCs are different, the cluster formed by their mergers  will inevitably have an internal metallicity spread \citep[e.g.,][]{mckenzie2021}.
Second, the star formation timescale could be longer than the SN timescale, if the GMCs are significantly more massive.
For example, in a cosmological RHD simulation of a dark matter halo with mass $\sim 5\times10^7\,\msun$ \citep{kimm2016}, the stellar metallicities of GC candidates increase up to $0.3\,Z_\odot$ from zero metallicity.
This is because the gas mass in the halo is an order of magnitude larger than the mass of the colliding GMCs and the star formation continues for $\sim 10\,{\rm Myr}$, allowing the SN ejecta to be recycled for $\Delta t\sim 6\,{\rm Myr}$.
Finally, as suggested in the literature \citep[e.g.,][]{kimjy2018}, multiple star formation episodes followed by metal enrichment on several hundreds of Myr-to-Gyr timescales may well explain the complex chemical abundance patterns observed in GCs \citep[][]{bastian2018,carretta2019}.
These diverse conditions will be the subject of future studies and our simulations should be regarded as controlled experiments for better understanding the detailed effects of radiation feedback in GC (candidate) formation.

\section{Discussion}

\subsection{Dominant radiation feedback process}

\begin{figure*}
\centering
\includegraphics[width=0.9\linewidth]{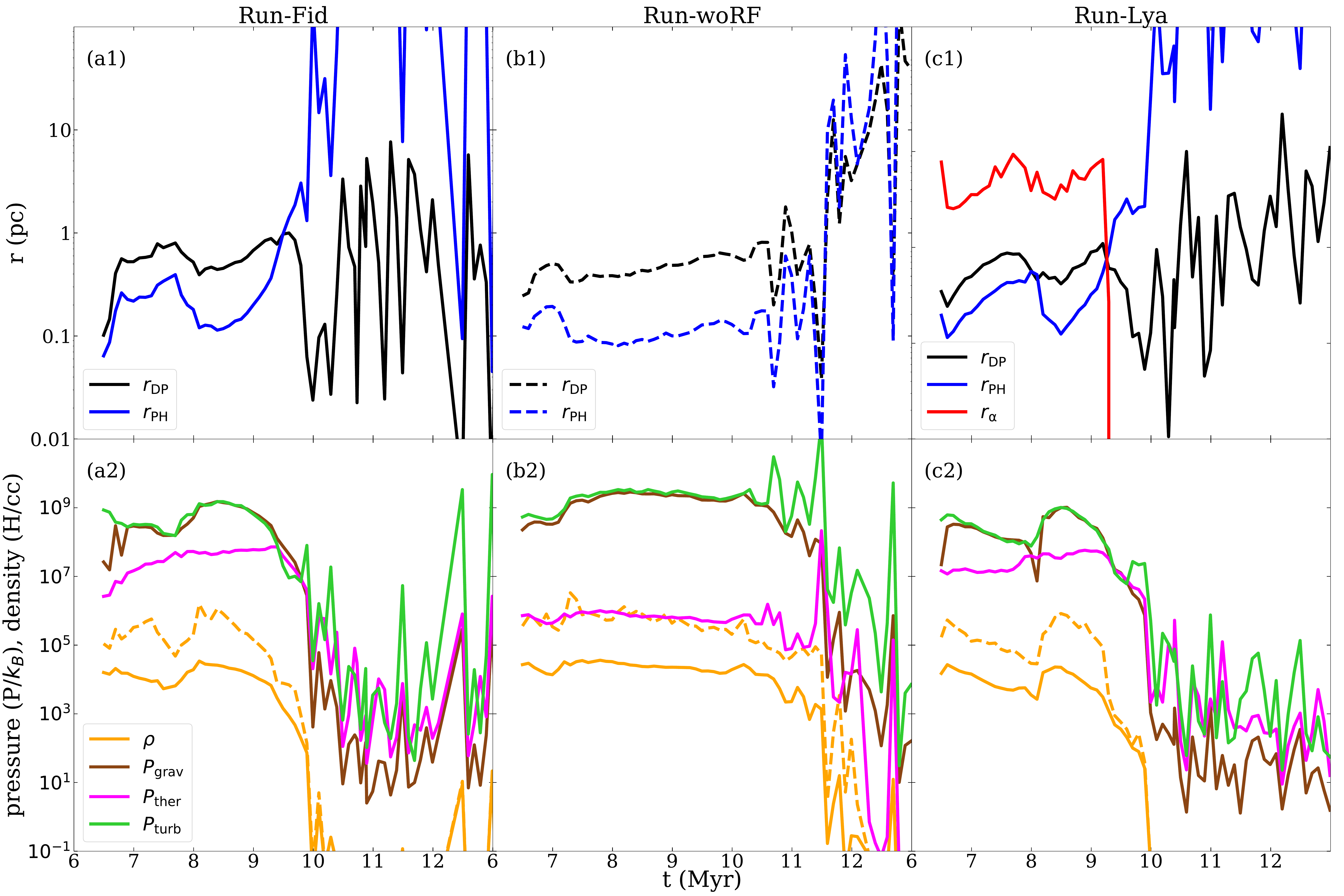}
\caption{The top panels display the maximum extent measured from the center of the potential GC to which photoionization heating ($r_{\rm PH}$, blue), direct radiation pressure ($r_{\rm DP}$, black), and \Lya\ pressure ($r_{\rm \alpha}$, red) can overcome the ISM pressure in three different runs. In \texttt{Run-woRF}, where the radiation feedback is absent, $r_{\rm PH}$ and $r_{\rm DP}$ are computed assuming that stellar particles radiate and are presented by dashed lines. The bottom panels exhibit volume-weighted pressure due to gravity (brown), thermal motion (magenta), and turbulence (green). The orange lines indicate the volume-weighted (solid) or mass-weighted (dashed) densities. All quantities are measured inside a sphere of 2 pc radius from the cluster center.}
\label{fig:r_fbk}
\end{figure*}

Several types of radiation feedback occur in GMCs, such as photoionization heating or direct radiation pressure.
To determine their relative importance in the suppression of star formation in GC-forming sites, we compute the maximum extent to which photoionization heating ($r_{\rm PH}$), direct radiation pressure ($r_{\rm DP}$), and \Lya\ pressure ($r_{\rm \alpha}$) balance the ISM pressure in Figure~\ref{fig:r_fbk}.
These scales are computed as \citep{rosdahl2015,kimm2017}
\begin{align}
r_{\rm PH} & = \left(\frac{3 N_{\rm ph, ion}}{4 \pi \alpha_{\rm B}}\right)^{1/3} \left( \frac{k_{\rm B} T_{\rm ion}}{P_{\rm ther} + P_{\rm grav} + P_{\rm turb}}\right)^{2/3}\\
r_{\rm DP} &= \sqrt{\frac{L_{\rm abs}}{4 \pi c \,(P_{\rm ther} + P_{\rm grav} + P_{\rm turb})}}\\
r_{\rm \alpha} & = \sqrt{\frac{M_F\,L_{\rm Ly\alpha}}{4 \pi c \,(P_{\rm ther} + P_{\rm grav} + P_{\rm turb})}},
\end{align}
where $N_{\rm ph,ion}$ is the number of ionizing photons radiated per unit time, $T_{\rm ion}=10^4\,{\rm K}$ is the temperature of the ionized medium, and $\alpha_B$ is the recombination coefficient at $T_{\rm ion}$ \citep{hui1997}.
The thermal and turbulent pressures are calculated within a sphere of radius 2 pc from the cluster center, as $P_{\rm ther} = \int (\rho_{\rm gas} / \mu m_{\rm H}) \, k_{\rm B}\, T dV / V$ and $P_{\rm turb} = \int \rho_{\rm gas} \,|\vec{v} - \left<\vec{v}\right>|^2 dV / V$, respectively, where $\mu$ is the mean molecular weight.
The pressure due to gravity is taken as $P_{\rm grav} = - \int \rho \Vec{g}\cdot\Vec{r} dV /V$.
To evaluate the effect of direct radiation pressure due to UV and optical photons, the luminosity absorbed by hydrogen, helium, and dust ($L_{\rm abs}$) is computed.
Additionally, the total Ly$\alpha$ luminosity emitted by gas within the same volume is computed from the recombinative and collisional radiation, as described in \citet{kimm2018}. 
The effect of non-thermal IR pressure is not presented here, as its contribution to the total pressure is negligible in our simulation.
This is because the typical optical depth of the IR photons by dust is small ($\tau_d \la 0.1$) and they simply escape without being trapped in such metal-poor environments.

Figure~\ref{fig:r_fbk} shows that in the fiducial run, direct radiation pressure more effectively controls gas collapse than photoionization heating while forming the potential GC population ($t\la9\,{\rm Myr}$).
This is also evident in \texttt{Run-woRF} if $r_{\rm PH}$ and $r_{\rm DP}$ are measured,  based on the luminosities expected from the stellar population (panel (b1)).
In the run with \Lya , $r_{\alpha}$ is even larger than $r_{\rm DP}$, indicating that \Lya\ pressure is the strongest form of radiation feedback early on.
However, $r_{\alpha}$ plunges at $t\ge9.2\,{\rm Myr}$ once the cloud is fully ionized and \Lya\ freely escapes the system.

Specifically, in the early stage of cluster formation ($6.5 \la t \la 7.5\,{\rm Myr}$), the stellar mass grows and the volume-weighted gas density within the central 2 pc sphere decreases (panels (a2) and (c2) in Figure~\ref{fig:r_fbk}).
Subsequently, the pressure due to gravity  and turbulence drops, which increases  $r_{\rm DP}$ and $r_{\rm PH}$ in the runs with radiation feedback. 
Note that the turbulent pressure and the pressure due to gravity are similar at $t\la 9\,{\rm Myr}$ (before the dense clumps are disrupted), as the initial virial parameters of the GMCs are set to $\alpha_{\rm vir}=0.7$.
Despite the decrease in density, the thermal pressure tends to build up or remain constant because the radiation feedback heats up the gas; however, $P_{\rm ther}$ is still subdominant compared to $P_{\rm grav}$ or $P_{\rm turb}$.
In contrast, $P_{\rm ther}$ in \texttt{Run-woRF} is smaller than that in the other runs by more than an order of magnitude, as the cloud temperature is kept cold ($T\sim 100\,{\rm K}$).
At $t \approx 8\,{\rm Myr}$, the gas density increases again in \texttt{Run-Fid} and \texttt{Run-Lya} due to the merger with other gas clumps, causing temporary reductions in both $r_{\rm DP}$ and $r_{\rm PH}$.
Once the direct radiation pressure removes the gas from the central region, photoionization takes over and $r_{\rm PH}$ rapidly expands at $t \approx 9\,{\rm Myr}$\footnote{We also confirm that the effect of photoelectric heating is subdominant compared to photoionization heating. This is done by simulating the ionization fronts in a uniform medium with $\nH=10^4\,\cmq$ and $T=100\,{\rm K}$ with and without photoelectric heating.}.
Then, the gas density in the cluster-forming region reduces to a level similar to the ISM ($n_{\rm H}\la 0.1\,\cmq$), and the absorption of optical and UV light becomes insignificant ($t \approx 10\,{\rm Myr}$). 

\begin{figure}
\centering
\includegraphics[width=\linewidth]{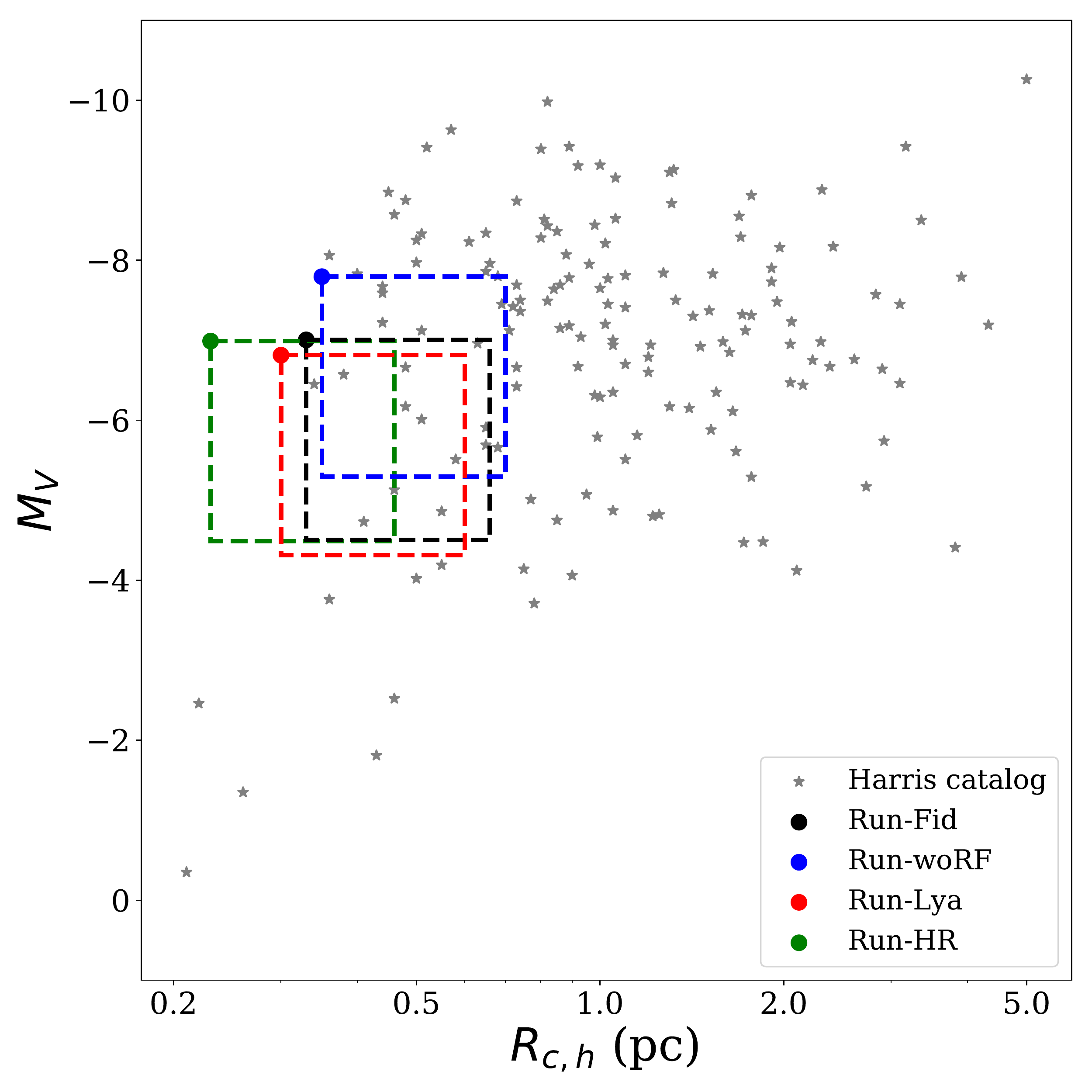}
\caption{Comparison of the V-band magnitude and half-mass radius ($R_{\rm c,h}$) between the Milky Way GCs (gray stars) and simulated GC candidates (colored circles). We adopt $M/L_V=2.0\,M_{\odot}/L_{V,\odot}$ to estimate the V-band magnitude of the simulated clusters, motivated by observational estimates \citep{baumgardt2017}. The dashed boxes denote the approximate range of the luminosity and radius evolution of the simulated clusters for a Hubble time, based on \citet{gieles2010,gieles2011}.}
\label{fig:rh_MV}
\end{figure}

\subsection{Compactness of the massive cluster}

Figures~\ref{fig:fig_fiducial} and \ref{fig:img_else} show that the formed clusters are generally very compact; the half-mass radius of the \texttt{Run-Fid} cluster is $R_{\rm c,h}=0.33\,{\rm pc}$ with a stellar mass of $M_{\rm GC}=1.083\times10^5\,\msun$ at the end of the simulation. The final mass of the cluster decreases by 16\% in \texttt{Run-Lya} or increases by a factor of $2.07$ in \texttt{Run-woRF}, but their $R_{\rm c,h}$ are still comparable to that of the fiducial case.
Furthermore, the cluster size is even more compact ($R_{\rm c,h}=0.23\,{\rm pc}$) in the run with the higher resolution (\texttt{Run-HR}), demonstrating that the size may not yet have converged.
This is likely because the turbulent structure of the cluster-forming region is not fully resolved.
As noted in previous studies, the Jeans length needs to be resolved using more than 32 cells to capture the turbulent structure in the ISM \citep[e.g.,][]{federrath2012}.
However, herein the Jeans length in the cluster forming region is often resolved using only 8 or so cells ($\lambda_J/\Delta x_{\rm min}\sim 1\,{\rm pc}/0.125\,{\rm pc}$).
Nevertheless, because the cluster is formed by mergers of gas clumps and stars bearing angular momentum, we believe that the resultant stellar structure will not be arbitrarily compact. 
Moreover, since the massive star cluster is better resolved in \texttt{Run-HR} than in the fiducial case, the cluster size is unlikely to dramatically change.
However, we note that the stellar mass of the cluster might be overestimated, given that an intermediate-mass black hole could form via stellar collisions in dense clusters \citep[e.g.,][]{katz2015}.

Figure~\ref{fig:rh_MV} compares the compactness of the simulated clusters (filled circles) with that of the present-day Milky Way GCs (gray points).
We employ the mass-to-light ratio of $M/L_V = 2.0\,M_{\odot}/L_{V,\odot}$, typical of GCs with $\sim\,10$ Gyr \citep{baumgardt2017}, to estimate the V-band luminosity of the simulated clusters.
We remind the reader that such cloud-cloud collisions may occur at high redshift ($z\ga 5$) during the formation of dwarf galaxies \citep[e.g.,][]{kimjh2018,sameie2022}.
Figure~\ref{fig:rh_MV} shows that at these luminosities, the simulated clusters are more compact than the Milky Way GCs ($0.2\la R_{\rm c,h}\la 5\,{\rm pc}$).
However, \citet{gieles2010, gieles2011} argued that the GCs are likely to lose up to $\sim\,90\%$ of their initial stellar mass due to stellar evolution, tidal disruption, and evaporation due to two-body relaxation, while their size increases by a factor of up to two \citep[see also][]{larsen2012, madrid2012, kruijssen2015,rodriguez2022}.
If these effects are considered, as shown by dashed boxes in Figure~\ref{fig:rh_MV}, it is possible that the simulated clusters evolve into normal present-day GCs.
Although a direct N-body simulation is needed to accurately follow the evolution of the GC candidate and to determine whether the structure can survive from tidal disruption, our results suggest that the cloud--cloud collision may be a viable route for the formation of GC.

\subsection{Caveats}

We present one of the first studies that attempt to understand the impact of radiation feedback on the formation of massive compact clusters using RHD \citep[see also,][]{dobbs2022}.
However, several caveats should be considered when interpreting our results. 
First, although a high resolution of up to $\Delta x_{\rm min} = 0.125\, {\rm pc}$ is employed herein, along with an aggressive refinement scheme to capture  turbulent interactions, the size of the simulated clusters does not fully converge.
As discussed in the previous section, the adoption of resolution elements that can resolve the Jeans length using more than 32 cells may produce a slightly more compact cluster. 
Second, our star particle of mass $100\,\msun$ represents a coeval cluster, but this should be modelled in a more realistic way by sampling the individual stellar mass for a given IMF \citep[e.g.,][]{geen2018}.
To mimic the uncertainty in the IMF sampling, additional simulations are performed with a larger stellar mass resolution of $m_{\rm star}^{\rm res}=1000\,\msun$, but the stellar mass of the potential GC is only changed by $\sim 20\%$ and our conclusions on the formation process remain largely unchanged. 
Third, several physical processes that may be relevant for GC formation are neglected.
\citet{grudic2022a} showed that although they are not responsible for disrupting clouds, protostellar jets can play an important role in regulating massive star formation.
Momentum-driven stellar winds may also disperse very dense gas in the vicinity of metal-rich O stars, although they are less effective than photoionization heating in destroying gas clouds with $\nH\sim 10^4$--$10^5 \,\cmq$ \citep[e.g.,][]{dale2013}.
These imply that the stellar mass in our simulations may be overestimated.
In contrast,  \citet{sakre2021} argued that strong magnetic fields delay gas collapse, increasing the number of massive dense cores when the field lines are aligned with the cloud collision axis.
Finally, the determination of the stellar mass formed during  GMC collisions may be affected by numerical diffusion.
As pointed out by \citet{pontzen2021}, numerical diffusion present in Eulerian codes tend to delay gas collapse.
This may increase the probability of the formation of a compact cluster due to the suppression of star formation in the moving GMC prior to the collision, although our higher resolution run, where diffusion is reduced somewhat demonstrates that this is not a dominant effect.
The inclusion of magnetic fields, which are believed to play such a role \citep{sakre2021} might also alleviate this numerical issue.

\section{Conclusions} \label{sec:discussion}

Motivated by recent numerical and observational findings \citep[e.g.,][]{kimjh2018,tsuge2021}, we simulated the formation of a massive compact cluster via the collision of two GMCs, each weighing  $3.6\times10^5\,\msun$, using RHD simulations.
In particular, we focused on the impact of radiation feedback on the formation process of GC candidates and found that it can play an important role in determining the star formation histories, stellar masses, and metallicity distributions \citep[see also][]{dobbs2022}. Our results are summarized as follows.

\begin{enumerate}

\item A massive star cluster ($\approx 10^5\,\msun$) is formed when two GMCs collide.
Roughly half of the cluster stars are formed in several gas clumps that collapse 1 Myr after the coalescence of the GMCs.
The other half is produced in the massive clump that is formed by the merging of the early star-forming clumps and the accretion of neighboring gas (Figure~\ref{fig:fig_fiducial}--\ref{fig:img_else}). In total, $\approx 15\%$ of the gas clouds turns into a GC candidate in 2--4 Myr.

\item Radiation feedback efficiently controls the growth of stellar mass in  cluster-forming gas clumps.
Star formation is completed in $\approx3\,{\rm Myr}$ in the presence of radiation feedback (\texttt{Run-Fid}), while the cluster-forming site keeps accreting the neighboring gas clumps and forming stars for $\approx 6\,{\rm Myr}$ when the radiation feedback is neglected (\texttt{Run-woRF}).
Thus, the cluster mass in \texttt{Run-woRF} is over-estimated by a factor of $\approx 2$, compared to that in the fiducial case  (Figure~\ref{fig:fig_SFH}).

\item Among the various radiation feedback processes considered  throughout the formation of the potential GC, the dominant mechanism for the suppression of the gas collapse is the direct radiation pressure, i.e., momentum transfer via the absorption of LyC radiation.
However, once the gas density drops to $\nH \la 10^3\, \cmq$ due to outflows driven by direct radiation pressure and gas consumption by star formation, photoionization is more dominant than direct radiation pressure in \texttt{Run-Fid} (Figure~\ref{fig:r_fbk}).
The impact of non-thermal radiation pressure due to IR photons is always negligible due to the low dust optical depth.

\item The additional pressure due to the multiple scatterings of \Lya\ photons partially inhibits the gas accretion onto the cluster-forming sites and suppresses the stellar mass by $16\%$ in \texttt{Run-Lya}.
The effect of Ly$\alpha$ pressure is only noticeable after a sufficient number of stars have formed and it progressively weakens as the star-forming region becomes ionized (Figure~\ref{fig:r_fbk}).

\item   The simulated clusters are much more compact than the present-day Milky Way GCs.
The half-mass radius of the cluster is $\approx 0.3\,{\rm pc}$ at the end of the simulations, which is about three times smaller than the typical size of Milky Way GCs.

\item Radiation feedback suppresses the internal metal enrichment due to SN explosions, which is consistent with the observations of a simple stellar population in the star clusters younger than $\sim2\,{\rm Gyr}$ \citep{bastian2018}.
However, 15\% of the cluster stars are self-enriched by SNe in the run without radiation feedback.

\end{enumerate}

Our numerical experiments focused on one way of forming a compact, massive star cluster during a head-on collision between two GMCs.
However, any process that triggers rapid gas collapse and star formation warrants investigation, including galaxy mergers \citep[e.g.,][]{whitmore1995,renaud2015,kimjh2018,lee2021}, to fully understand the GC formation process.
Although the star cluster mass may be overestimated herein due to the absence of other forms of feedback, such as protostellar jets or stellar winds, the impact of direct radiation pressure and photoionization is likely to be persistent in other GC formation scenarios.
As shown in Section~\ref{sec:metal}, the early feedback affects the local environment into which SNe propagate, and hence the metal enrichment in the star-forming clouds and neighboring medium.
This implies that chemical mixing is complex in star-forming sites and warrants further studies on the detailed chemical evolution of GCs \citep[e.g.,][]{tenorio2015,bastian2018,kimjy2018,calura2019}. 

\acknowledgments

TK was supported by the National Research Foundation of Korea (NRF) grant funded by the Korea government (No. 2020R1C1C1007079 and No. 2022R1A6A1A03053472), and acted as the corresponding author.
The supercomputing time for numerical simulations was kindly provided by KISTI (KSC-2019-CRE-0196), and large data transfer was supported by KREONET, which is managed and operated by KISTI. This work was also performed using the DiRAC Data Intensive service at Leicester, operated by the University of Leicester IT Services, which forms part of the STFC DiRAC HPC Facility (www.dirac.ac.uk). The equipment was funded by BEIS capital funding via STFC capital grants ST/K000373/1 and ST/R002363/1 and STFC DiRAC Operations grant ST/R001014/1. DiRAC is part of the National e-Infrastructure.

\bibliographystyle{aasjournal}
\bibliography{references}

\end{document}